\documentclass[lettersize,journal]{IEEEtran}

\usepackage{color}
\usepackage{graphicx, subfigure}
\usepackage{amsmath}
\usepackage{cite}
\usepackage{amsfonts}
\usepackage{multirow}
\usepackage{url}
\usepackage{threeparttable}
\usepackage{rotating}
\usepackage{hyperref}

\begin{document}
 
\title{Semantic-aware Digital Twin for\\ AI-based CSI Acquisition}
\author{\normalsize {Jiajia~Guo, \IEEEmembership{\normalsize {Member,~IEEE}},
Yiming~Cui, \IEEEmembership{\normalsize {Graduate Student Member,~IEEE}},
and Shi Jin, \IEEEmembership{\normalsize {Fellow,~IEEE}}
}

}

\maketitle

\begin{abstract}
Artificial intelligence (AI) substantially enhances channel state information (CSI) acquisition performance but is limited by its reliance on single-modality information and deployment challenges, particularly in dataset collection.
This paper investigates the use of semantic-aware digital twin (DT) to enhance AI-based CSI acquisition. 
We first briefly introduce the motivation and recent advancements in AI-driven CSI acquisition and semantic-aware DT employment for air interfaces. Then, we thoroughly explore how semantic-aware DT can bolster AI-based CSI acquisition. We categorizes the semantic-aware DT for AI-based CSI acquisition into two classes: enhancing AI-based CSI acquisition through integration with DT and using DT to aid AI-based CSI deployment. Potential integration frameworks are introduced in detail.
Finally, we conclude by outlining potential research directions within the semantic-aware DT-assisted AI-based CSI acquisition.
\end{abstract}
 
\IEEEpeerreviewmaketitle

\section{Introduction}

Massive multiple-input multiple-output (MIMO), by equipping the base station (BS) with a large number of antennas, serves as the key enabling technology of 5G.
In alignment with the dramatically heightened requirements of 6G \cite{ITU2030}, the massive MIMO technology will be further enhanced and evolved \cite{union2022future}.
Antenna numbers will markedly rise to hundreds or thousands (defined as extreme MIMO in \cite{union2022future}), significantly boosting spectrum and energy efficiency, network coverage, etc., while innovative variants like holographic MIMO are also emerging.
The potentials of massive MIMO and its variants are contingent upon the high-quality optimization (e.g., precoding design) on the basis of channel state information (CSI).
However, due to the high antenna dimension of massive MIMO, the acquisition of high-quality CSI results in a significant overhead
and high computational complexity, which are intolerable for practical systems.
Consequently, developing an efficient technology for CSI acquisition is crucial to harness the full potential of massive MIMO in 6G.

Artificial intelligence (AI), which is capable of automatically learning knowledge from large datasets, has been introduced to wireless communications.
AI native has been acknowledged as a pivotal feature of the 6G air interface \cite{union2022future}, and also been adopted in CSI acquisition tasks, such as channel estimation \cite{8640815}, feedback \cite{guo2022overview}, and prediction \cite{9277535}.
The integration of AI into CSI acquisition offers two primary benefits, i.e., performance improvement (including accuracy and overhead) and complexity reduction.
Nonetheless, there are two principal drawbacks prevalent in most existing studies. Firstly, existing studies tend to concentrate on enhancing CSI acquisition performance through the deployment of more advanced AI technologies, while disregarding valuable knowledge from other domains/modals. Furthermore, the integration of AI into CSI acquisition brings novel challenges associated with practical deployment that are often overlooked.

Digital twin (DT) \cite{9854866}, which replicates the physical world into a digital virtual world,
is poised to be a pivotal application and service in 6G, as highlighted in \cite{ITU2030} and termed as \emph{wireless for DT}.
Meanwhile, DT also holds the potential to enhance wireless communications, termed as \emph{DT for wireless}.
The iterative trial-and-error approaches to radio access network (RAN) deployment and optimization are resource-intensive. DT technology can substantially reduce these expenses by offering a virtual model of the RAN for comprehensive testing and optimization.
Numerous manufacturers have implemented DT-based RAN deployment and optimization in practical systems, including Huawei's IntelligentRAN and NVIDIA's Omniverse \cite{10148936}.
Additionally, there have been efforts to incorporate DT into the air interface, such as the beam prediction \cite{10283592}.
Despite this, there is a lack of extensive research focusing on CSI acquisition, with only a few studies, such as \cite{9897088,240219434}, addressing this fine-grained task.
Nevertheless, as previously discussed, the high overhead and limited accuracy within CSI acquisition have posed significant challenges to fully realizing the potential of massive MIMO.
Furthermore, existing works aim to construct a DT replica that closely mirrors the real-world environment, utilizing all available information to achieve this, and then employ it to assist CSI acquisition. However, this approach leads to considerable inefficiency.
Given these challenges, it is crucial to integrate efficient DT into AI-based CSI acquisition.

Herein, we primarily concentrate on utilizing  semantic-aware DT for AI-based CSI acquisition.
Specifically, we first briefly introduce the motivation and recent progress of AI for CSI acquisition and DT for air interface, respectively.
Subsequently, we conduct a thorough investigation into how semantic-aware DT can be leveraged to enhance AI-based CSI acquisition.
We categorize the approaches into two classes: \emph{the integration of AI and semantic-aware DT for CSI acquisition} and \emph{the use of semantic-aware DT to facilitate the deployment of AI-based CSI acquisition}.
Finally, we highlight some potential research directions in semantic-aware DT for AI-based CSI acquisition.

\section{Motivation and Recent Progress}
\label{s2}
This section introduces the motivation and recent progress of AI for CSI acquisition and DT for air interface, respectively.

\subsection{AI for CSI acquisition}

The main challenges in traditional CSI acquisition involve acquisition performance (accuracy and overhead) and computational complexity.
To this end, AI has been thoroughly integrated into the CSI acquisition framework, spanning pilot design, channel estimation, and feedback \cite{guo2022overview}.

Unlike traditional algorithms that often assume a certain channel distribution first and rely heavily on manually designed features, neural networks (NNs) in AI can automatically extract features based on the environment knowledge (i.e., channel distributions) learned from training datasets \cite{guo2025prompt}.
Therefore, the CSI acquisition NNs trained with a large number of CSI samples can perform well on the CSI dataset with the same channel distributions, thereby enhancing CSI acquisition accuracy and reducing CSI acquisition overhead.
In contrast to conventional iterative algorithms, the inference process of AI-based CSI acquisition can be substantially expedited with the assistance of graphics processing units (GPUs).

As mentioned before, AI-based CSI acquisition faces two main challenges:
\textbf{1). Integration and utilization of multimodal information:}  CSI are indicators of the propagation environment, which also can be characterized by the information in other modalities, such as vision and point cloud data. The integration and utilization of this new information can improve the performance of AI-based CSI acquisition, warranting in-depth investigation and exploration.
\textbf{2). Practical deployment challenges in AI-based CSI acquisition:} AI-driven CSI acquisition fundamentally relies on substantial data during both the pre-training and online learning phases. This presents significant challenges, introducing considerable overhead to practical systems—a factor not considered in traditional air interface studies.
 
\subsection{DT for air interface}
\label{DTforAir}
 DT in wireless is usually regarded as a replica of the physical propagation environment.
 The decisions (e.g., beam management \cite{10283592}) in real wireless systems can be performed according to the knowledge obtained from the digital virtual world.
 Consequently, the DT for air interface encompasses two principal components: the creation of a replica for the propagation environment (i.e., a digital virtual world) and its corresponding utilization strategy to support decision-making in real systems.

\begin{figure*}[t]
    \centering
    \includegraphics[width=0.7\textwidth]{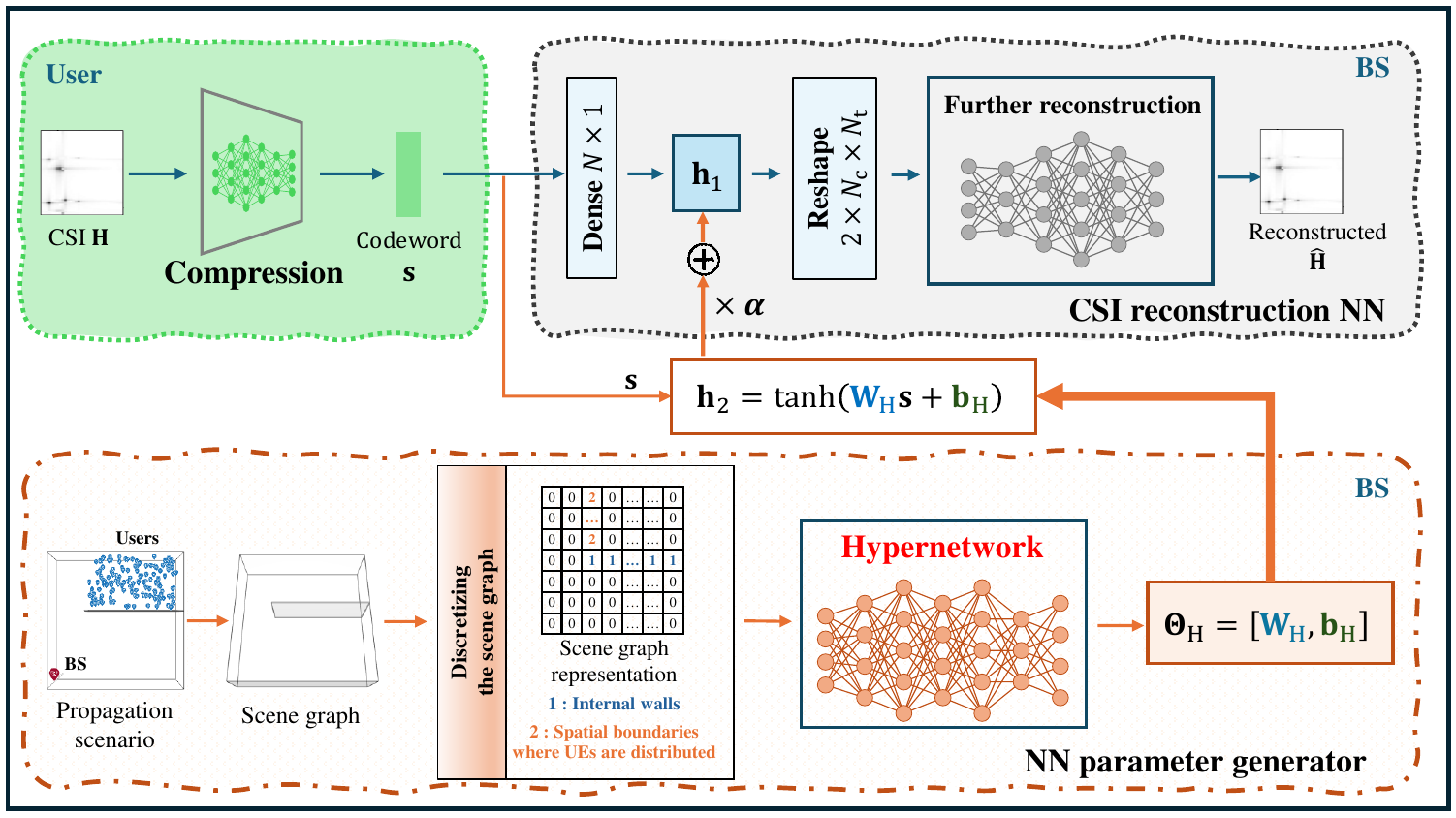}
    \caption{\label{HypernetworkImag}A semi-black box approach for AI-based CSI feedback, as proposed in \cite{liu2025adapcsinet}. The hypernetwork processes the discretized scene graph to produce the weights and biases of the personalized dense layer, which then processes the feedback codeword to enable scene-specific CSI reconstruction.}
\vspace{-0.6cm}
\end{figure*}

\begin{figure}[t]
    \centering
    \includegraphics[width=0.45\textwidth]{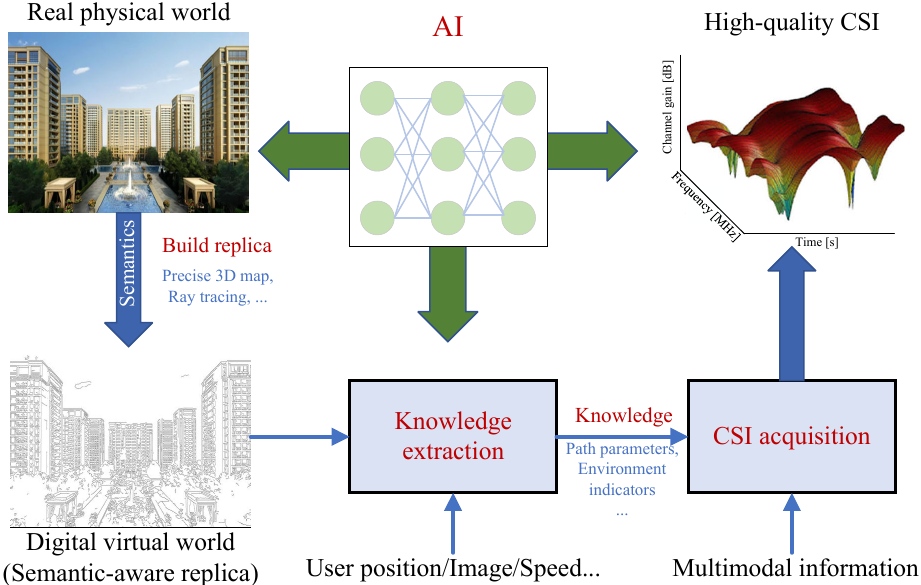}  \caption{\label{AI&DT1}Illustration of integration of AI and semantic-aware DT for enhanced CSI acquisition. Semantic-aware DT acts as a fresh knowledge/information source while AI integrates into various modules, facilitating KE from DT and the acquisition of CSI.}
    \vspace{-0.6cm}
\end{figure}

 \textbf{How to efficiently build a replica of the physical propagation environment?}
With the rapid advancements in three-dimensional (3D) mapping technology, including equipment for data collection and processing, it is now possible to generate detailed 3D maps, such as NVIDIA's Omniverse \cite{10148936}. These maps include precise positioning, materials, and other relevant details for both communication equipment and surrounding objects (such as trees and roads) within the propagation environment \cite{10198573}.
Another widely used tool for replica building is ray tracing simulator, exemplified by commercial software like Wireless InSite and open-source platforms such as Sionna \cite{aoudia2025sionna}.
The ray tracing simulator can generate the corresponding channel response (including each path information) according to the propagation environment, generated by 3D computer graphics software tools (e.g., Blender), with a given communication setting.
Due to Sionna's open-source accessibility, ray tracing simulators have become a popular research tool in wireless communications, bridging digital worlds and complex channels effectively.
{However, constructing a DT replica that fully mirrors the real-world environment using all available information is highly inefficient and often impractical. Instead, only the components affecting channel distribution are essential.  For instance, in \cite{liu2025adapcsinet}, a scene graph is employed to create a DT that efficiently captures the transmission environment. We refer to this streamlined, efficient DT as a semantic-aware DT, which captures environmental semantics critical to channel characteristics while maintaining high efficiency.
}

{For instance, Figure \ref{HypernetworkImag} depicts an AI-based CSI feedback framework enabled by a semantic-aware DT in \cite{liu2025adapcsinet}, with the creation process shown in the bottom left part. In a simplified indoor scenario, the primary environmental features affecting propagation-such as wall locations, dimensions, and room layout-are selected. These are discretized into a matrix, where walls and open spaces are represented by distinct values. This discretized matrix can be regarded as a semantic-aware DT and significantly enhances CSI feedback accuracy. }

\begin{figure*}[t]
    \centering
    \includegraphics[width=0.75\textwidth]{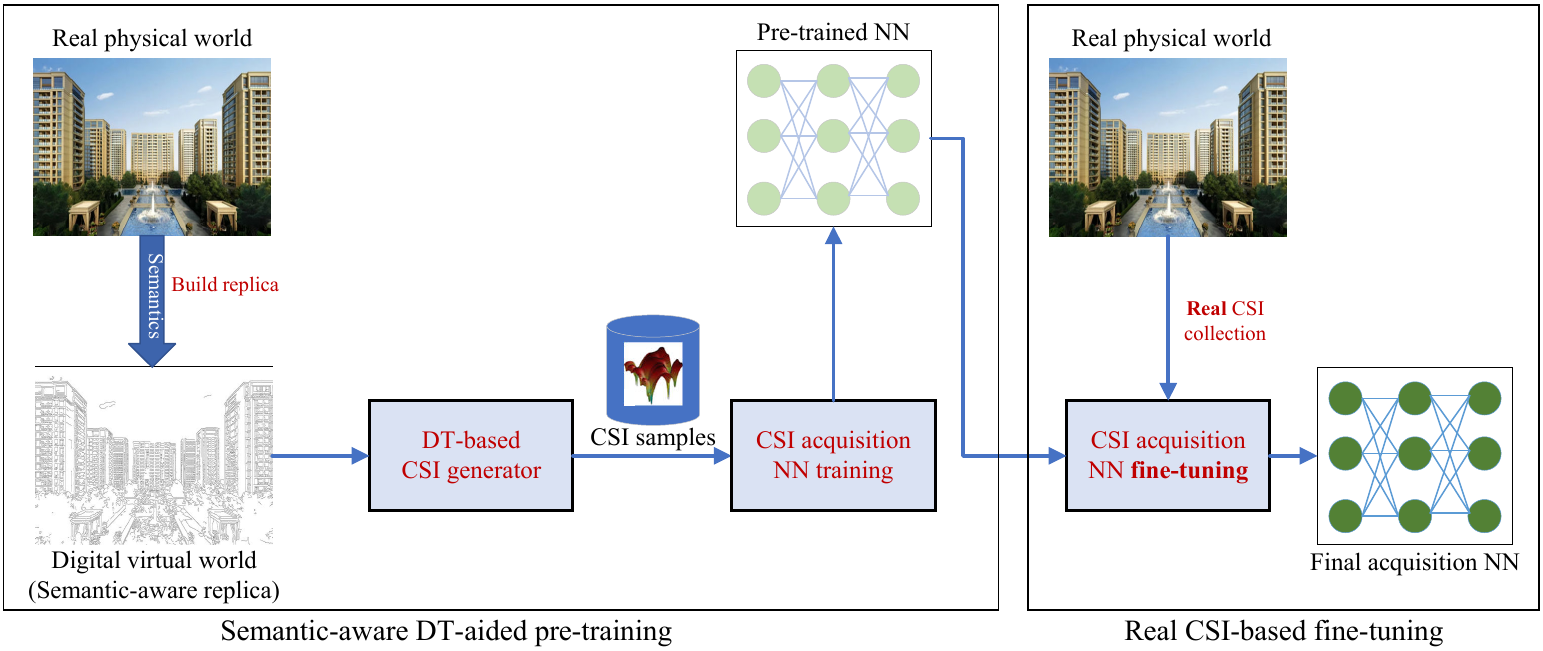}    \caption{\label{Deployment1}Illustration of deploying AI-based CSI acquisition through semantic-aware DT: semantic-aware-DT-based white-box training data generation. The training process is structured as a two-stage methodology, where the initial phase employs CSI data generated by the semantic-aware DT for the preliminary NN pre-training, and the subsequent phase leverages practical system-collected CSI to fine-tune NN.}
    \vspace{-0.6cm}
\end{figure*}

 \textbf{How to enhance decision-making in real systems with a replica?}
 Some communication tasks, such as beam selection, heavily relies on the environment around the user. If the user position is available with a global navigation satellite system (GNSS) or a camera, the surrounding environment can be properly assessed through a precise 3D map, enabling the selection of the optimal beam for effective communication.
 Meanwhile, a ray tracing simulator can provide valuable channel path information through real-time simulations, aiding in the design and optimization of communication systems.
However, there have been limited studies \cite{9897088,240219434} that focus on enhancing or supporting AI-based CSI acquisition through DT.

\section{Semantic-aware DT for AI-based CSI Acquisition}

CSI acquisition is not only a fundamental aspect but also the core module underlying wireless communications. AI has demonstrated considerable potential in enhancing CSI acquisition's performance. This section will comprehensively explore the utilization of semantic-aware DT for AI-based CSI acquisition, which falls into two distinct categories: integration of AI and semantic-aware DT for enhanced CSI acquisition, and deployment of AI-based CSI acquisition utilizing semantic-aware DT. The former category is focused on enhancing the performance benefits achieved through AI integration, whereas the latter aims to address the challenges associated with the deployment of AI-based CSI acquisition using semantic-aware DT.

\subsection{Integration of AI and semantic-aware DT for enhanced CSI acquisition}

AI, as a data-driven technology, is frequently employed as a tool to either partially or totally replace conventional algorithms, delivering enhancements in accuracy and efficiency.
In contrast, DT is not a tool but rather a new dimension that provides extra information and has not been previously investigated.
In this part, we will elucidate the overall architecture of AI and semantic-aware DT integration for enhanced CSI acquisition before delineating the key modules involved.

\subsubsection{Overall Framework}

Figure \ref{AI&DT1} provides the illustration of integration of AI and semantic-aware DT for enhanced CSI acquisition. Semantic-aware DT acts as a fresh source of knowledge/information while AI integrates into various modules, facilitating knowledge extraction (KE) from DT and the acquisition of CSI.
More specifically, the CSI acquisition, augmented by semantic-aware DT assistance, encompasses three principal stages, throughout which AI is intricately woven.
The detailed stages are as follows:
\begin{itemize}
    \item The first stage, the same as the first problem mentioned in Section \ref{DTforAir}, is to build a semantic-aware replica of the real physical world. 
    \item Subsequently, in the second stage, named ``knowledge extraction'', the module extracts pertinent knowledge and information for CSI acquisition. This extraction is based on the semantic-aware digital replica constructed during the initial stage, in conjunction with real-world user data.
    \item In the final stage, CSI is estimated or reconstructed using the knowledge extracted in the second phase, along with some real-world multimodal data, such as received pilot signals in channel estimation, historical CSI in channel prediction, and user positional and visual information.
\end{itemize}

In summary, the initial stage is centered on constructing an accurate digital replica of the physical propagation environment; whereas the last stage is dedicated to the AI-based CSI acquisition based on multimodal data, which includes inputs from the real propagation world as well as knowledge offered by the KE.
Since the article focuses on exploring the DT's role in improving CSI acquisition with AI, the initial and final phases are not the primary focus of this discussion.
Readers interested in the two stages can refer to \cite{10198573} for details on the creation of a replica and to \cite{9277535} for an in-depth exploration of multimodal learning applied to CSI acquisition.
The following part will provide a detailed introduction to the KE module implemented in the second stage.

\subsubsection{DT-aided Knowledge Extraction for CSI Acquisition}

The second stage, that is, the KE module, poses two primary questions: what information is transmitted to the KE module, and what output this module can generate.
\emph{The key point here lies in the function of DT in enhancing the performance of CSI acquisition.}

\textbf{Inputs to KE Module:}
The inputs to the KE module are designed to facilitate comprehension of the user's surrounding propagation environment, that affects the signal transmission, within the replica world, thereby enriching the information pool for CSI acquisition.
CSI acquisition falls into two categories: instantaneous, which captures current channel conditions, and predictive, which forecasts future channel states.
Each category requires different inputs for the KE module.

\emph{Instantaneous CSI acquisition:} In a stable environment, the characteristic of a user's CSI is influenced by its location. With accurate location data, we can model its signal propagation in the replica environment through techniques such as ray tracing, thereby obtaining the high-quality CSI characteristic. Consequently, the challenge becomes how to accurately determine the user's position in the real world. A vanilla approach to acquire the user's position is to use the built-in localization system, such as GNSS. Nonetheless, GNSS may not always be available to the user, and even with GNSS, localization precision might not meet the requirements. Alternative techniques, like vision-based localization, must be employed to guarantee high-accuracy and stable positioning.

\emph{Future CSI acquisition:} Considering channel aging and the potential of eliminating the overhead of CSI acquisition, CSI prediction, producing the future CSI according to the historical information, stands out as a promising approach.
Consequently, the inputs to KE module should be helpful in facilitating the trend acquisition within the propagation environment.
The user's historical positions, movement direction, and speed, which can be obtained by several sensors (e.g., inertial measurement unit) and techniques (e.g., GNSS), can be input to help predict the user trend.

\textbf{Outputs of KE Module:}
The outputs of the KE module demonstrate what information the semantic-aware DT can supply to aid AI-based CSI acquisition.
We classify this information into the following two main categories.

\emph{Information with direct correlation to CSI:} The CSI matrix characterizes the signal transmission within a specific propagation environment, taking into account the multipath effects where the signal propagates through various paths. Within a simulated replica, certain path parameters (e.g., path loss, angle of arrival, and angle of departure) can be accurately determined. However, stochastic factors, such as system noise and non-idealities of transceivers, avoid exact measurement and might not be detectable at all. Nevertheless, the path information obtained from the semantic-aware replica can substantially enhance the performance of CSI acquisition. For example, with the such information's aid, more effective pilot signals  can be designed, and the process of channel estimation and feedback can be improved through the use of this additional/side information.

\emph{Information with indirect correlation to CSI:} The information that is useful to CSI acquisition protocol design also needs to be extracted. Acquiring CSI is a complex procedure that requires meticulous design beyond just the detailed acquisition algorithms. The semantic-aware DT replica can offer assistance across multiple parts apart from the detailed algorithms. The acquisition of CSI relies on the support of a comprehensive protocol,  including the specification of the density, structure, and type of pilot signals, the time intervals for performing estimation and feedback, etc. Extracting related information from the DT replica and incorporating it into protocol design is highly beneficial. For instance, if slow channel changes within a stable propagation environment are observed in the semantic-aware DT replica, larger channel estimation and feedback intervals can be adopted, thereby substantially mitigating CSI acquisition overhead. 
 
\begin{figure}[t]
    \centering
    \includegraphics[width=0.45\textwidth]{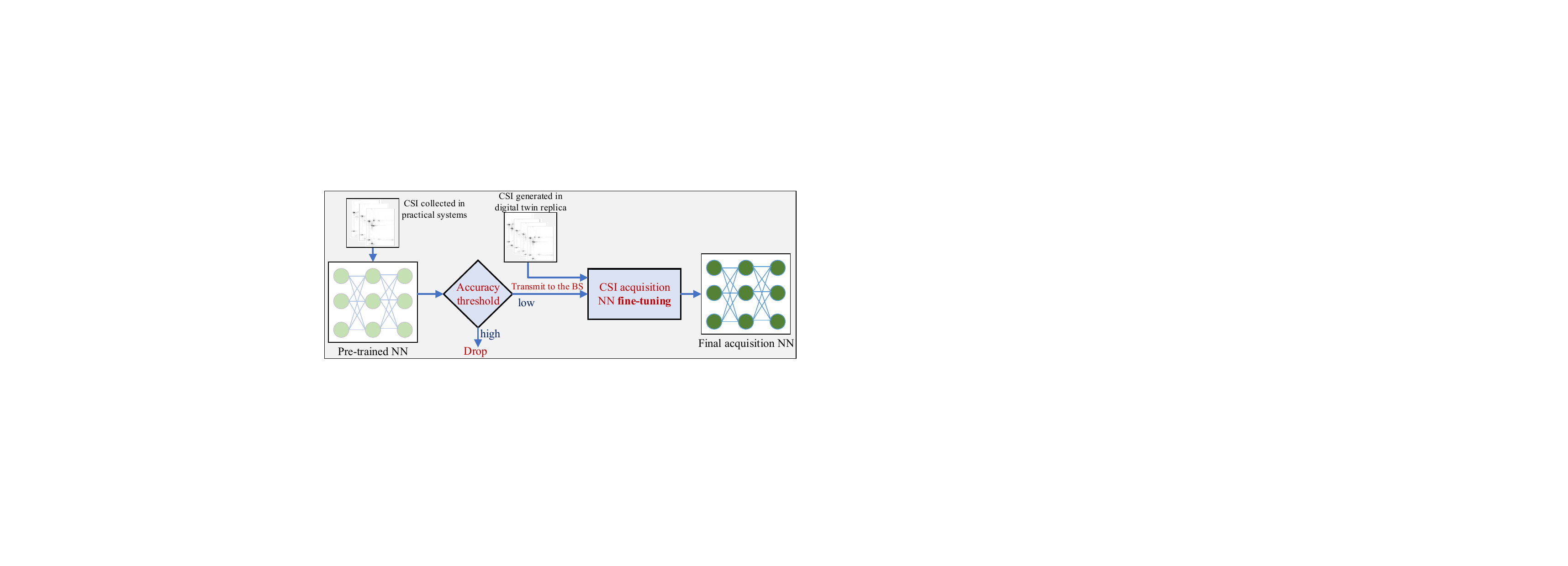}    \caption{\label{sampleSelection}Illustration of the refinement data selection proposed in \cite{240219434}. 
CSI samples exhibiting low acquisition performance with the NNs pre-trained using DT data should be selected for further fine-tuning of the NNs alongside the DT data.}
\vspace{-0.4cm}
\end{figure}

\begin{figure}[t]
    \centering    \includegraphics[width=0.45\textwidth]{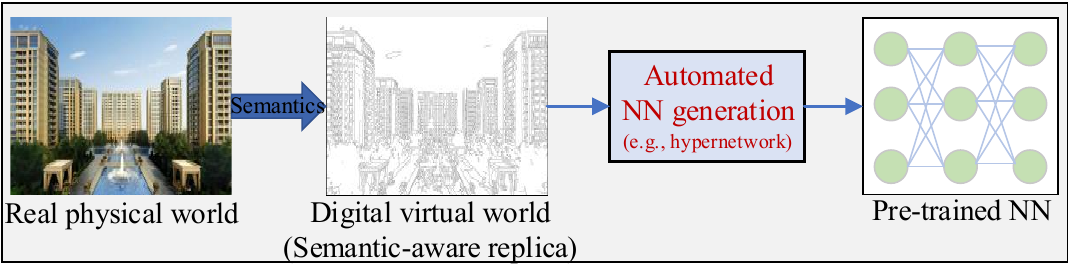}
    \caption{\label{blackBox}Illustration of semantic-aware-DT-based black-box training data generation.}
    \vspace{-0.6cm}
\end{figure}

\subsection{Deployment of AI-based CSI acquisition using DT}

While AI integration substantially boosts CSI acquisition, it concurrently presents novel challenges when deployed. 
AI-based CSI acquisition, since it's completely reliant on data, first has to overcome the issue of data during deployment.
In AI-based CSI acquisition, data is needed for two key phases: the NN training prior to deployment and the online NN training to address generalization issues.
Both before and during deployment, collecting training data poses significant challenges, including data collection overhead and time constraints.
In this subsection, we discuss two semantic-aware DT-based approaches—white-box methods, which generate explicit training data, and black-box methods, which directly produce NN parameters—to address these challenges.

\begin{figure*}[t]
    \centering
    \includegraphics[width=0.7\textwidth]{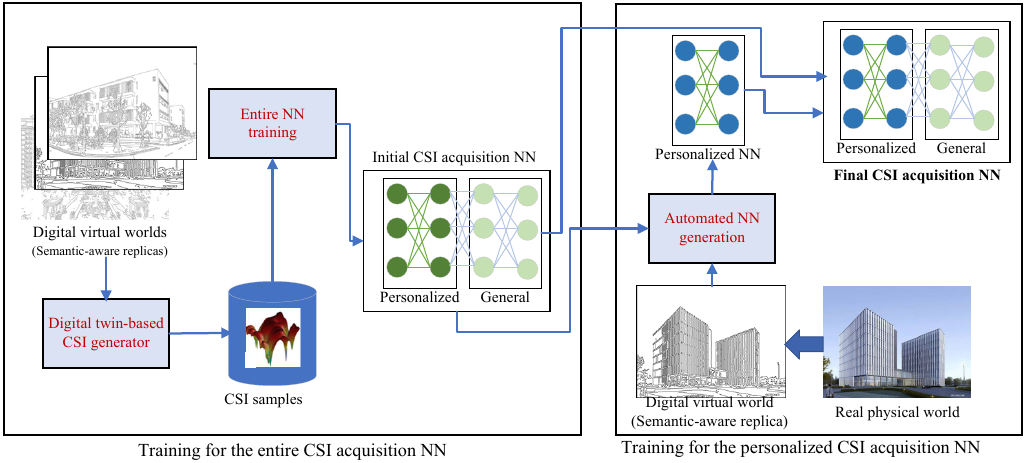}    \caption{\label{semiBlackBox}Semantic-aware-DT-based semi-black box training data generation, which entails segmenting the entire CSI acquisition NN into two components: a general NN and a personalized NN. The general NN is trained using samples from diverse semantic-aware DT worlds, whereas the personalized NN is fine-tuned with data from a specific semantic-aware DT world.}
        \vspace{-0.6cm}
\end{figure*}

\subsubsection{White-box Training Data Generation based on Semantic-aware DT}
It is a logical approach to employ semantic-aware DT to assist in generating CSI training samples, thereby reducing the overhead caused by data collection.
Figure \ref{Deployment1} illustrates the training framework of the CSI acquisition NN, augmented by a semantic-aware DT-based white-box training data generation method. The NN training is partitioned into two distinct phases:

\textbf{Pre-training with Data Generated by semantic-aware DT:}
The training of the NN in AI-based CSI acquisition involves learning the channel distributions, and then, based on these learned distributions, designing automatic algorithms to perform channel estimation, feedback, and other operations.
The channel distribution, determined by the physical environment, is reflected in the channel dataset from which the NN learns to infer and adapt its parameters for accurate modeling.
{Given that he semantic-aware DT mirrors real-world CSI distribution,}
the channel distribution of the CSI within the semantic-aware DT replica closely parallels that in the real propagation environment.
Therefore, it is a natural idea to utilize the CSI samples generated (collected) within the DT replica as a dataset for training the CSI acquisition NNs \cite{240219434}.

Once the efficient semantic-aware DT replica is established, CSI samples can be acquired using methods like ray tracing, with user positions being randomly assigned.
However, this method presents two main problems as follows,
\begin{itemize}
    \item \emph{Unknown user position distribution:} For simplification, most existing studies typically posit users within a regularly shaped service region, such as a rectangle or circle. For instance, \cite{240219434} assigns user locations randomly within a $200m \times 230m$ rectangular zone. Nevertheless, user position distribution tends to be irregular in practical systems. The distribution of user positions greatly influences the channel distribution within a specific cell, making it crucial to ascertain potential user positions. In outdoor environments, most users carry GNSS-enabled devices, facilitating the easy acquisition of user position data. Conversely, indoor user position distribution is challenging to ascertain and requires advanced methodologies.
    \item \emph{High complexity of semantic-aware DT-based CSI generation:} The CSI generation process demands substantial computational resources. Although GPUs can accelerate this process, CSI generation remains a protracted activity. Two possible solutions exist to address this issue. The first is simplification of the propagation parameters involved.
    For instance, limiting the number of signal paths could markedly diminish the complexity of the CSI generation process. Alternatively, reducing the quantity of CSI that needs to be generated is another solution. As long as the generated CSI samples can accurately represent the channel distribution, there is no necessity to produce additional CSI samples any more. During the training phase, data augmentation techniques can be utilized to enhance NN performance.
\end{itemize}

\textbf{Fine-tuning with Data collected in Practical Systems:} {Unavoidable mismatches between the real world and the simplified semantic-aware DT replica result in discrepancies in channel distributions. First, the DT replica cannot fully capture the real world \cite{240219434}, as unpredictable factors may influence signal propagation, and accurately describing real-world semantics is challenging. Consequently, the semantic-aware DT prioritizes efficiency at the expense of accuracy. Second, as previously noted, precisely determining the user position distribution—a critical factor affecting channel distribution—poses a significant challenge in practical systems. Therefore, fine-tuning the CSI acquisition NN, pre-trained with semantic-aware DT data, is essential to mitigate these discrepancies and enhance performance.}

However, this extra fine-tuning of the CSI acquisition NN requires CSI samples collected from practical systems, resulting in a considerable amount of data collection overhead.
To mitigate this overhead, it is not necessary to collect all the CSI samples. Instead, focus can be placed on gathering only those samples that are beneficial for the fine-tuning process and are not present in the dataset generated by the semantic-aware DT replica.
{For example, the DT replica in \cite{240219434} assumes the same geometric layout as the real world but omits objects like trees, and accordingly, \cite{240219434} proposes a data selection-based fine-tuning framework, as shown in Figure \ref{sampleSelection}, to mitigate this overhead.} Specifically, only the CSI samples exhibiting low acquisition performance with the NNs pre-trained using DT data should be selected for further fine-tuning.
While this can significantly enhance CSI acquisition accuracy and diminish the data collection burden in practical systems, there remains a question as to whether the CSI samples that perform poorly with the pre-trained NNs are indeed missing from the DT data.
As demonstrated in \cite[Figure 34]{guo2022overview}, a subset of samples exhibits low acquisition accuracy despite being included in the training dataset.
Hence, a superior data collection algorithm is needed.

\subsubsection{Black-box Training Data Generation based on Semantic-aware DT}

The discussed white-box approach adheres to the conventional framework of data-driven NN training.
The DT replica aids in the deployment of CSI acquisition NNs by providing training data that follows a distribution similar to practical systems.
However, this method still occupies a large amount of computational resources to pre-train CSI acquisition NNs.
Exploring the use of a semantic-aware DT to completely eliminate the pre-training process is worthwhile.

The essence of CSI acquisition NN training is to learn the channel distribution from training samples. Therefore, in the white-box method, data generated by the DT replica is utilized to help NNs learn the channel distribution. This distribution is determined by the propagation environment and the distribution of user positions. In other words, the channel distribution has a mapping relationship with both the propagation environment and the user position distribution.
Given a pre-defined NN architecture, the NN parameters are determined by both the propagation environment and the user position distribution, which are assumed to be integrated within the semantic-aware DT world in this article.
Therefore, it makes logical sense to directly generate the parameters of the CSI acquisition NNs from the semantic-aware DT world.
We consider this approach as semantic-aware DT-based black-box training data generation, which implicitly generates data.

Figure \ref{blackBox} illustrates the framework for semantic-aware-DT-based black-box training data generation framework, utilized for the NN parameter generation of pre-trained CSI acquisition NNs. Initially, a semantic-aware DT world is constructed, mirroring the real propagation world, and encompasses comprehensive details such as the propagation environment and user position distribution.
The DT does not need to fully capture all information from the real world but should include key semantics, such as buildings and their materials, that influence signal transmission.
Subsequently, such a semantic-aware DT world is fed into an NN parameter generation module to produce the pre-trained NN parameters.
The primary challenge lies in the NN parameter generation module that generates NN parameters based on the semantic-aware DT world. This module encompasses three main issues:
\begin{itemize}
    \item \emph{Input form:} The complex DT world can be represented in the form of various modalities, such as precise 3D map and OBJ files in Blender. Assuming the NN generation module is NN-based, its input must be in the form of matrix or tensor. Therefore, how to translate the DT world into a matrix or tensor formats. In the new format, the input should selectively preserve features that impact channel distribution, streamlining the data while ensuring accurate channel distribution. 
    \item \emph{Module design: }The purpose of this module is to produce NN parameters, usually accomplished with the assistance of a hypernetwork \cite{ha2017hypernetworks}.
    The fundamental concept behind the hypernetwork is to produce NN parameters based on its input. For instance, in order to enhance the robustness of NN-based channel estimation against varying signal-to-noise ratios (SNR), the hypernetwork generares the parameters for the channel estimation NNs using the SNR as the input variable. Similarly, the DT world represented in new formats can be input into the hypernetwork, which in turn generates the parameters for the CSI acquisition NNs.
    \item \emph{Module complexity: }The parameter number of CSI acquisition NNs is typically quite vast, which inevitably contributes to significant complexity within the NN parameter generation module. During the online updating of NN parameters for generalizability improvement, this disadvantage might present an obstacle to its deployment.
    Therefore, it is essential to reduce the burden of the NN parameter generation module.
\end{itemize}

Considering the challenges mentioned, employing a purely black-box method with considerable complexity poses serious hurdles and may be infeasible in practical systems.
A semi-black box strategy could be more viable.
Figure \ref{semiBlackBox} illustrates the core structure of a semi-black box approach for CSI acquisition. This approach divides the CSI acquisition NN into two components: a general NN and a personalized NN. The general NN is trained using samples from diverse DT worlds, while the personalized NN is fine-tuned with data from a specific DT world. Thus, the training process involves two key phases. Initially, the entire CSI acquisition NN, comprising both general and personalized components, is trained using CSI samples from various DT environments. Subsequently, a specific semantic-aware DT world is input into the automated NN generation module to produce parameters for the personalized NN, enabling tailored adaptation to the target environment.

{
Figure \ref{HypernetworkImag} illustrates a semi-black box approach for AI-based CSI feedback, as proposed in \cite{liu2025adapcsinet}. Upon receiving the feedback codeword, the BS reconstructs the downlink CSI using NNs in two stages: an initial reconstruction via a dense layer and a subsequent refinement using a complex NN. 
In the first training phase, a general reconstruction NN, encompassing both the dense layer and the complex NN, is trained with CSI samples derived from diverse semantic-aware DT worlds, represented as scene graphs. Subsequently, a hypernetwork generates an additional dense layer, serving as the personalized NN, by taking the scene graph of a specific environment as input. Specifically, the hypernetwork processes the discretized scene graph to produce the weights and biases of the personalized dense layer, which then processes the feedback codeword to enable scene-specific CSI reconstruction.}
Given that the automated NN generation module doesn't have to produce all NN parameters and that the entire NNs have been pre-trained, this substantially lessens the complexity and workload of the module. Furthermore, while the pre-trained NN from the first phase might not reach the performance of a tailored NN for the scenario, it may still perform satisfactorily.

\section{Conclusion and Potential Research Directions}
{This paper explores the way for efficient CSI acquisition in 6G systems by leveraging semantic-aware DTs, offering a foundation for future advancements of massive MIMO.}
Semantic-aware DTs for AI-based CSI acquisition involves the integration of AI and semantic-aware DT for CSI acquisition and the use of semantic-aware DT to facilitate the deployment of AI-based CSI acquisition. The potential frameworks are introduced in detail.
However, this field still has several important challenges and areas for further study:
\begin{itemize}
    \item \emph{Semantic-aware DT world (replica) creation: }The discussions above presupposes the precise creation of a semantic-aware DT world (replica), and an imprecise replica would significantly compromise the proposed DT-assisted strategies. Accurately replicating real-world environment wireless semantics and user positional data poses significant challenges and sometimes may even be impractical. Therefore, semantic-aware-DT-assisted strategies can now initially be studied and applied in simpler, more controllable environments, like that of an intelligent factory.
    \item \emph{Multimodal information utilization: }A fundamental attribute of semantic-aware-DT-assisted AI in CSI acquisition involves leveraging multimodal information. Nevertheless, the methodologies for integrating this multimodal data remain unclear and acquire further investigation. AI is considered a promising solution for addressing this complex challenge.
    \item \emph{User position distribution modeling: }The channel distribution is influenced not just by the propagation environment but also by the distribution of user positions. To the best of our knowledge, there has been little exploration into learning user position distribution. This issue transcends traditional wireless communication challenges and necessitates interdisciplinary research.
    \item \emph{Collaboration of multiple DTs: }In scenarios such as cell-free networks, there is a requirement to acquire the CSI of a user from various access points. This necessitates collaboration among multiple DT worlds to ensure high-quality acquisition of CSI.
    \item \emph{Standardization: }Unlike the transceiver design, the CSI acquisition requires strict standardization. The introduction of a DT world presents significant challenges. The standardization need necessitate as much flexibility as possible for BSs and users while ensuring performance.
    \item \emph{Privacy protection: }For the purpose of efficient CSI acquisition, the DT world requires gathering user behavioral data, including but not limited to location and speed, presenting a risk of privacy leaks. Consequently, there is a need to develop specialized privacy protection mechanisms, which only collects wireless environment semantics.
\end{itemize}
\bibliographystyle{IEEEtran}
\bibliography{magazine}

\begin{thebibliography}{10}
\providecommand{\url}[1]{#1}
\csname url@samestyle\endcsname
\providecommand{\newblock}{\relax}
\providecommand{\bibinfo}[2]{#2}
\providecommand{\BIBentrySTDinterwordspacing}{\spaceskip=0pt\relax}
\providecommand{\BIBentryALTinterwordstretchfactor}{4}
\providecommand{\BIBentryALTinterwordspacing}{\spaceskip=\fontdimen2\font plus
\BIBentryALTinterwordstretchfactor\fontdimen3\font minus \fontdimen4\font\relax}
\providecommand{\BIBforeignlanguage}[2]{{%
\expandafter\ifx\csname l@#1\endcsname\relax
\typeout{** WARNING: IEEEtran.bst: No hyphenation pattern has been}%
\typeout{** loaded for the language `#1'. Using the pattern for}%
\typeout{** the default language instead.}%
\else
\language=\csname l@#1\endcsname
\fi
#2}}
\providecommand{\BIBdecl}{\relax}
\BIBdecl

\bibitem{ITU2030}
\BIBentryALTinterwordspacing
{ITU-R WP5D}, ``Framework and overall objectives of the future development of {IMT} for 2030 and beyond,'' Tech. Rep., Jun. 2023, {A}ccessed on Mar. 20, 2025. [Online]. Available: \url{https://www.itu.int/md/R19-WP5D-230612-TD-0905/en}
\BIBentrySTDinterwordspacing

\bibitem{union2022future}
\BIBentryALTinterwordspacing
{ITU-R M.2516-0}, ``Future technology trends of terrestrial international mobile telecommunications systems towards 2030 and beyond,'' Tech. Rep., Nov. 2022, {A}ccessed on Mar. 20, 2025. [Online]. Available: \url{https://www.itu.int/dms_pub/itu-r/opb/rep/R-REP-M.2516-2022-PDF-E.pdf}
\BIBentrySTDinterwordspacing

\bibitem{8640815}
M.~Soltani, V.~Pourahmadi, A.~Mirzaei, and H.~Sheikhzadeh, ``Deep learning-based channel estimation,'' \emph{IEEE Commun. Lett.}, vol.~23, no.~4, pp. 652--655, Apr. 2019.

\bibitem{guo2022overview}
J.~Guo, C.-K. Wen, S.~Jin, and G.~Y. Li, ``Overview of deep learning-based {CSI} feedback in massive {MIMO} systems,'' \emph{IEEE Trans. Commun.}, vol.~70, no.~12, pp. 8017--8045, Dec. 2022.

\bibitem{9277535}
Y.~Yang, F.~Gao, C.~Xing, J.~An, and A.~Alkhateeb, ``Deep multimodal learning: Merging sensory data for massive {MIMO} channel prediction,'' \emph{IEEE J. Sel. Areas Commun.}, vol.~39, no.~7, pp. 1885--1898, Jul. 2021.

\bibitem{9854866}
L.~U. Khan, Z.~Han, W.~Saad, E.~Hossain, M.~Guizani, and C.~S. Hong, ``Digital twin of wireless systems: Overview, taxonomy, challenges, and opportunities,'' \emph{IEEE Commun. Surveys Tuts.}, vol.~24, no.~4, pp. 2230--2254, 4th Quart. 2022.

\bibitem{10148936}
X.~Lin, L.~Kundu, C.~Dick, E.~Obiodu, T.~Mostak, and M.~Flaxman, ``6{G} digital twin networks: From theory to practice,'' \emph{IEEE Commun. Mag.}, vol.~61, no.~11, pp. 72--78, Nov. 2023.

\bibitem{10283592}
S.~Jiang and A.~Alkhateeb, ``Digital twin based beam prediction: Can we train in the digital world and deploy in reality?'' in \emph{Proc. IEEE Int. Conf. Commun. Workshops (ICC Workshops)}, 2023, pp. 36--41.

\bibitem{9897088}
B.~Vilas~Boas, W.~Zirwas, and M.~Haardt, ``Machine learning for {CSI} recreation in the digital twin based on prior knowledge,'' \emph{IEEE Open J. Commun. Soc.}, vol.~3, pp. 1578--1591, 2022.

\bibitem{240219434}
S.~Jiang and A.~Alkhateeb, ``Digital twin aided massive {MIMO: CSI} compression and feedback,'' in \emph{Proc. IEEE Int. Conf. Commun. (ICC)}, 2024, pp. 1--6.

\bibitem{guo2025prompt}
J.~Guo, Y.~Cui, C.-K. Wen, and S.~Jin, ``Prompt-enabled large {AI} models for {CSI} feedback,'' \emph{arXiv preprint arXiv:2501.10629}, 2025.

\bibitem{liu2025adapcsinet}
J.~Liu, J.~Guo, Y.~Cui, C.-K. Wen, and S.~Jin, ``{AdapCsiNet}: Environment-adaptive {CSI} feedback via scene graph-aided deep learning,'' \emph{arXiv preprint arXiv:2504.10798}, 2025.

\bibitem{10198573}
A.~Alkhateeb, S.~Jiang, and G.~Charan, ``Real-time digital twins: Vision and research directions for {6G} and beyond,'' \emph{IEEE Commun. Mag.}, vol.~61, no.~11, pp. 128--134, Nov. 2023.

\bibitem{aoudia2025sionna}
F.~A. Aoudia, J.~Hoydis, M.~Nimier-David, S.~Cammerer, and A.~Keller, ``{Sionna RT}: Technical report,'' \emph{arXiv preprint arXiv:2504.21719}, 2025.

\bibitem{ha2017hypernetworks}
D.~Ha, A.~M. Dai, and Q.~V. Le, ``Hypernetworks,'' in \emph{Proc. 5th Int. Conf. Learn. Representations (ICLR)}, 2017, pp. 1--18.

\end{thebibliography}

\end{document}